\documentclass[useAMS,usenatbib]{mn2e} 
\usepackage{aas_macros}
\usepackage{graphics}
\usepackage{times}
\usepackage{amsmath,amssymb}
\usepackage{bm}
\usepackage{color}
\usepackage{hyperref}

\def\be{\begin{equation}}
\def\ee{\end{equation}}
\def\ba{\begin{eqnarray}}
\def\ea{\end{eqnarray}}

\def\lsim{\mathrel{\rlap{\lower4pt\hbox{\hskip1pt$\sim$}}
    \raise1pt\hbox{$<$}}}                
\def\gsim{\mathrel{\rlap{\lower4pt\hbox{\hskip1pt$\sim$}}
    \raise1pt\hbox{$>$}}}

\pagerange{\pageref{firstpage}--\pageref{lastpage}}
\pubyear{2015}

\begin{document}

\label{firstpage}

\title[Angle-only estimators and the GREAT3 simulations]{A demonstration of position angle-only weak lensing shear estimators on the GREAT3 simulations}
\author[Lee Whittaker, Michael L. Brown \& Richard A. Battye]{Lee Whittaker,
  Michael L. Brown, \& Richard A. Battye\\ Jodrell Bank Centre for Astrophysics,
  School of Physics and Astronomy, University of Manchester, Oxford
  Road, Manchester M13 9PL} \date{\today}

\maketitle

\begin{abstract}
We develop and apply the position angle-only shear estimator of \cite{whittaker14} to realistic galaxy images. This is done by demonstrating the method on the simulations of the GREAT3 challenge \citep{mandelbaum14}, which include contributions from anisotropic PSFs. We measure the position angles of the galaxies using three distinct methods - the integrated light method, quadrupole moments of surface brightness, and using model-based ellipticity measurements provided by {\tt IM3SHAPE}. A weighting scheme is adopted to address biases in the position angle measurements which arise in the presence of an anisotropic PSF. Biases on the shear estimates, due to measurement errors on the position angles and correlations between the measurement errors and the true position angles, are corrected for using simulated galaxy images and an iterative procedure. The properties of the simulations are estimated using the deep field images provided as part of the challenge. A method is developed to match the distributions of galaxy fluxes and half-light radii from the deep fields to the corresponding distributions in the field of interest. We recover angle-only shear estimates with a performance close to current well-established model and moments-based methods for all three angle measurement techniques. The Q-values for all three methods are found to be $Q\sim400$. The code is freely available online at \url{http://www.jb.man.ac.uk/~mbrown/angle_only_shear/}.
\end{abstract}


\section{Introduction}
\label{sec:intro}
Gravitational lensing provides a way of probing the dark Universe. As light travels to us from distant objects, it is deflected by foreground mass inhomogeneities. This effect can be exploited to provide information on the distribution of dark matter and the nature of dark energy, enhancing our understanding of the late-time evolution of the Universe. In the weak regime, the lensing of light by the large-scale structure of the Universe induces small but coherent distortions of observed background galaxy shapes. Since one does not know the intrinsic galaxy shapes, statistical methods must be employed to extract information about the underlying mass distribution. 

If we assume that the size of the source object is much smaller than the angular scale over which the gravitational potential of the foreground object changes, the distortion of the image can be described by a linearized mapping of the object from the source plane to the image plane (see \cite{bartelmann01} for an in-depth discussion). This linear transformation can be separated into two components; the convergence, which magnifies the image, and the shear, which both magnifies and changes the shape of the image. In this paper, we are concerned with estimating the shear from a field of simulated galaxy images. Assuming that the shape of a source galaxy can be described by an ellipse, the shear changes the ellipticity of the galaxy in the image plane. If one assumes that the distribution of the intrinsic ellipticities (i.e. the ellipticities in the source plane) are randomly orientated, then an accurate measurement of an individual galaxy's ellipticity provides a noisy, unbiased estimate of the shear at the position of the galaxy. This is the standard method for performing a weak lensing analysis. Such an approach has already been used to constrain various cosmological parameters, such as the amplitude of the matter power spectrum (e.g. \citealt{brown03, hoekstra06, fu08}) and the dark energy equation of state (e.g. \citealt{schrabback10, kilbinger13}).

In order to perform weak lensing using the standard method, one requires accurate measurements of the galaxy shapes. There are generally two classes of shape measurement techniques \citep{viola14}; these are the moments-based and the model-based methods. Moments-based methods measure the ellipticity of a galaxy by calculating the quadrupole moments of the light distribution of the pixelized galaxy image. Examples of such algorithms include the Kaiser-Squires-Broadhurst (KSB) method \citep{kaiser95} and the DEIMOS method \citep{melchior11}. Model-based methods fit a parameterized galaxy model to a pixelized galaxy image by finding the extremum of a loss function. Examples of this approach include {\tt \emph{lens}fit} (\citealt{miller07,kitching08}) and {\tt IM3SHAPE} \citep{zuntz13}.

For any realistic weak lensing survey, measurements of the galaxy shapes need to be recovered from noisy, pixelized images. These images are convolved with an instrumental and/or atmospheric point spread function (PSF), and both moments and model-based methods must correct for the PSF contribution. Moments based methods generally also implement a weighting function to reduce the effects of noise at large scales, and this function must subsequently be corrected for. For the case of an isotropic PSF, an incorrect calibration of the required correction will lead to multiplicative biases in the shear estimates. However, the orientation of a galaxy will remain unaltered. The same is true if one multiplies the galaxy image by a circular weighting function centered on the centroid of the galaxy. In \cite{whittaker14} (hereafter W14), we proposed a method of performing weak lensing using only measurements of the galaxy orientations with the aim of exploiting this property. We demonstrated the method using simple simulations where we ignored the effects of a PSF and considered a simple S\'ersic galaxy model for the intrinsic galaxy shapes. We found that the position angle-only method has the potential to yield shear estimates with a performance comparable with the KSB method.

An anisotropic PSF will bias position angle measurements and, therefore, estimates of the shear if not corrected for. This bias is addressed in this paper where we build upon the ideas introduced in W14 and apply the angle-only method to the simulations of the control-ground-constant ({\tt cgc}) branch of the GREAT3 simulations \citep{mandelbaum14}. These simulations were designed to test the performance of a shear estimator on realistic galaxy images and include the effects of noise, pixelization, an anisotropic PSF, and realistic distributions of galaxy flux, size and shape.

In Section \ref{sec:a_o_method}, we summarize the angle-only method proposed in W14 and discuss how noise on the position angle measurements impacts the analysis. We describe the three methods used to measure the position angles of the galaxies in Section \ref{sec:ang_meas}. The application of the angle-only method to the GREAT3 simulations is presented in Section \ref{sec:g3_sims}, where we describe the weighting scheme used to mitigate the effects of an anisotropic PSF and the calibration simulations implemented to remove the effects of the biasing introduced by measurement errors on the position angles. The results are presented in Section \ref{sec:results}, where we compare the angle-only method, using the three angle measurement techniques, with the results of a naive application of {\tt IM3SHAPE} and also with the highest entries to the GREAT3 challenge from {\tt IM3SHAPE} and the KSB method. We conclude with a discussion in Section \ref{sec:discussion}.

\section{The angle-only method}
\label{sec:a_o_method}
In this section, we briefly describe the angle-only shear estimator introduced in W14.

Working within the regime of weak lensing, we can express the observed complex ellipticity of a galaxy, $\bm{\epsilon}^{\mathrm{obs}}$, in terms of the intrinsic ellipticity, $\bm{\epsilon}^{\mathrm{int}}$, and the reduced shear, $\bm{g}$:
\begin{equation}\label{eq:observed_e}
\bm{\epsilon}^{\mathrm{obs}}=\frac{\bm{\epsilon}^{\mathrm{int}}+\bm{g}}{1+\bm{g}^{*}\bm{\epsilon}^{\mathrm{int}}},
\end{equation}
where the asterisk denotes complex conjugation. We can also express $\bm{\epsilon}^{\mathrm{obs}}$ in polar form, such that
\begin{align}\label{eq:eobs_polar}
\epsilon_1^{\mathrm{obs}}=&\left|\bm{\epsilon}^{\mathrm{obs}}\right|\cos\left(2\alpha\right),\nonumber\\
\epsilon_2^{\mathrm{obs}}=&\left|\bm{\epsilon}^{\mathrm{obs}}\right|\sin\left(2\alpha\right),
\end{align}
where $\alpha$ is the observed position angle of the galaxy. If we assume a constant shear, and that the intrinsic ellipticities are drawn randomly from an azimuthally symmetric probability distribution, $f\left(\bm{\epsilon}^{\mathrm{int}}\right)$, the means of the cosines and sines of the position angles can be written as
\begin{align}\label{eq:mean_cos_sin}
\left<\cos\left(2\alpha\right)\right>=&F_1\left(\left|\bm{g}\right|\right)\cos\left(2\alpha_0\right),\nonumber\\
\left<\sin\left(2\alpha\right)\right>=&F_1\left(\left|\bm{g}\right|\right)\sin\left(2\alpha_0\right),
\end{align}
where $\alpha_0$ is the shear position angle. The function $F_1\left(\left|\bm{g}\right|\right)$ is found to be
\begin{align}\label{eq:general_F}
F_1\left(\left|\bm{g}\right|\right)=\, &\frac{1}{\pi}\int_0^{\left|\bm{\epsilon}_{\mathrm{max}}^{\mathrm{int}}\right|}\int_{-\frac{\pi}{2}}^{\frac{\pi}{2}}\mathrm{d}\alpha^{\mathrm{int}}\mathrm{d}\left|\bm{\epsilon}^{\mathrm{int}}\right|f\left(\left|\bm{\epsilon}^{\mathrm{int}}\right|\right)\nonumber\\
&\times h_1\left(\left|\bm{g}\right|,\left|\bm{\epsilon}^{\mathrm{int}}\right|,\alpha^{\mathrm{int}}\right),
\end{align}
where $\alpha^{\rm int}$ denotes the intrinsic position angle. The function $h_1\left(\left|\bm{g}\right|,\left|\bm{\epsilon}^{\mathrm{int}}\right|,\alpha^{\mathrm{int}}\right)$ is given as
\begin{equation}\label{eq:g_function}
h_1\left(\left|\bm{g}\right|,\left|\bm{\epsilon}^{\mathrm{int}}\right|,\alpha^{\mathrm{int}}\right)=
\frac{E_1}{\sqrt{E_1^2+E_2^2}},
\end{equation}
with
\begin{align}\label{eq:e'}
E_1=&\left|\bm{g}\right|\left(1+\left|\bm{\epsilon}^{\mathrm{int}}\right|^2\right)+\left(1+\left|\bm{g}\right|^2\right)\left|\bm{\epsilon}^{\mathrm{int}}\right|\cos\left(2\alpha^{\mathrm{int}}\right),\nonumber\\
E_2=&\left(1-\left|\bm{g}\right|^2\right)\left|\bm{\epsilon}^{\mathrm{int}}\right|\sin\left(2\alpha^{\mathrm{int}}\right).
\end{align}
From here, W14 showed that, for a finite number of source galaxies, an estimate of the orientation of the shear is given by
\begin{equation}\label{eq:est_ao}
\hat{\alpha}_0=\frac{1}{2}\tan^{-1}\left(\frac{\sum_{i=1}^N\sin(2\alpha^{(i)})}{\sum_{i=1}^N\cos(2\alpha^{(i)})}\right),
\end{equation}
where $N$ is the number of galaxies in the sample. A preferred value of the $F_1\left(\left|\bm{g}\right|\right)$ function can also be recovered for the sample using
\begin{equation}\label{eq:value_F}
F_1\left(\left|\hat{\bm{g}}\right|\right)=\sqrt{\left[\frac{1}{N}\sum_{i=1}^N\cos\left(2\alpha^{(i)}\right)\right]^2+\left[\frac{1}{N}\sum_{i=1}^N\sin\left(2\alpha^{(i)}\right)\right]^2}.
\end{equation}
This relation can then be inverted to provide an estimate of $\left|\bm{g}\right|$. We can combine this estimate with an estimate of $\hat{\alpha}_0$ provided by equation (\ref{eq:est_ao}) to recover a complete estimate of the shear.

Measurement errors on the position angles bias estimates of the mean trigonometric functions in equations (\ref{eq:est_ao}) and (\ref{eq:value_F}). This bias propagates into estimates of the shear. As discussed in W14, for a general error distribution we can write the measured position angle as
\begin{equation}\label{eq:alpha_err}
\hat{\alpha}=\alpha+\delta\alpha,
\end{equation}
where $\delta\alpha$ is the error on the measurement. The means of the trigonometric functions are then
\begin{align}\label{eq:trig_estimates_full}
\left<\cos\left(2\hat{\alpha}\right)\right>=&\left<\cos\left(2\alpha+2\delta\alpha\right)\right>,\nonumber\\
\left<\sin\left(2\hat{\alpha}\right)\right>=&\left<\sin\left(2\alpha+2\delta\alpha\right)\right>.
\end{align}
We can write equation (\ref{eq:trig_estimates_full}) in terms of the covariance between the true position angles and the measurement errors, such that
\begin{align}\label{eq:cov_c_s_relation}
\left<\cos\left(2\hat{\alpha}\right)\right>=&C'+\mathrm{cov}\left(\cos\left(2\alpha\right),\cos\left(2\delta\alpha\right)\right)\nonumber\\
&-\mathrm{cov}\left(\sin\left(2\alpha\right),\sin\left(2\delta\alpha\right)\right),\nonumber\\
\left<\sin\left(2\hat{\alpha}\right)\right>=&S'+\mathrm{cov}\left(\sin\left(2\alpha\right),\cos\left(2\delta\alpha\right)\right)\nonumber\\
&+\mathrm{cov}\left(\cos\left(2\alpha\right),\sin\left(2\delta\alpha\right)\right),
\end{align}
where we define
\begin{align}\label{eq:cov_c_s}
C'=&\left<\cos\left(2\alpha\right)\right>\beta_c-\left<\sin\left(2\alpha\right)\right>\beta_s,\nonumber\\
S'=&\left<\sin\left(2\alpha\right)\right>\beta_c+\left<\cos\left(2\alpha\right)\right>\beta_s,\nonumber\\
\beta_c=&\left<\cos\left(2\delta\alpha\right)\right>,\nonumber\\
\beta_s=&\left<\sin\left(2\delta\alpha\right)\right>.
\end{align}
Hence, we see that errors on the position angles of the galaxies and correlations between the true lensed position angles and the measurement errors bias our estimates of the mean trigonometric functions. This in turn biases the shear estimates. W14 addressed this issue, in the absence of a PSF, by estimating the $\beta$ and covariance terms using simulations and applying an iterative procedure. However, we find that this approach is inadequate for fields which have a large anisotropic PSF. In Section \ref{sec:g3_sims}, we introduce an alternative iterative procedure. We use an initial estimate of the shear recovered using weighted averages of the trigonometric functions to simulate the expected bias in the shear estimates and subsequently correct our estimates.

\section{Measuring the position angles}
\label{sec:ang_meas}
When analysing the {\tt cgc} branch of the GREAT3 simulations, we use three distinct methods to measure the position angles from the simulated galaxy images. The first is the integrated light method discussed in W14. The second uses the second order moments of brightness of the galaxy image. The third uses ellipticities measured by {\tt IM3SHAPE} to determine the position angles. Here we briefly discuss the three approaches.

\subsection{The integrated light method}
\label{subsec:int_light}
The integrated light method for measuring galaxy position angles is discussed in detail in W14. Here we discuss the application of the method to the images of the {\tt cgc} simulations.

We begin by estimating the mean half-light radius of the galaxies in the field of interest using the relationship between the mean half-light radius and the mean flux of the galaxies discussed in Section \ref{sec:g3_sims} and given in equation (\ref{eq:mean_hlr}). For each galaxy in the {\tt cgc} branch, we initially assume the centroid of the galaxy to be the centre of the image. We apply a circular Gaussian weighting function to the image centered on this initial estimate of the centroid. The half-light radius of the weighting function is equal to twice the mean half-light radius of the galaxies in the field. The image is then convolved with a circular Gaussian kernel with a width of two pixels to reduce the effects of pixelization. The centroid is then recalculated using the first-order moments of the convolved weighted surface brightness distribution, $I_w(\bm{\theta})$, with the moments defined as 
\begin{equation}\label{eq:moments}
Q_{ij...k}=\frac{\int\mathrm{d}^2\theta I_w(\bm{\theta})\theta_i\theta_j...\theta_k}{\int\mathrm{d}^2\theta I_w(\bm{\theta})},
\end{equation}
and where $\bm{\theta}$ is the angular position of the galaxy image on the sky. This step is iterated until the difference between subsequent estimates of the components of the centroid are less than $10^{-3}$ of a pixel.

We then estimate the 1D integrated light distribution, $I'\left(\theta\right)$, of the galaxy as a function of assumed position angle, $\theta$, by integrating $I'(\bm{\theta})$ over the radial component centered on the estimated centroid, 
\begin{equation}\label{eq:brightness_marg}
I'(\theta)=\int\mathrm{d}r I_w(r,\theta).
\end{equation}
The details of how this integral is performed are discussed in W14. Finally, the estimated position angle of the galaxy is given by
\begin{equation}\label{eq:alpha_est_dist}
\hat{\alpha}=\frac{1}{2}\tan^{-1}\left(\frac{\int\mathrm{d}\theta I'\left(\theta\right)\sin\left(2\theta\right)}{\int\mathrm{d}\theta I'\left(\theta\right)\cos\left(2\theta\right)}\right).
\end{equation}

\subsection{A moments based method}
\label{subsec:moments}
The moments based method uses second order moments of the convolved weighted brightness distribution, $I_w(\bm{\theta})$, to estimate the position angle \citep{kaiser95}. We follow the approach outlined in the previous subsection to estimate the width of the weighting function used for each field and the centroid of the individual galaxy images. The position angles of the galaxies are then calculated using the second-order moments of $I_w(\bm{\theta})$, as defined in equation (\ref{eq:moments}), such that
\begin{equation}\label{eq:moments_angles}
\hat{\alpha}=\frac{1}{2}\tan^{-1}\left(\frac{2Q_{11}}{Q_{20}-Q_{02}}\right).
\end{equation}

\subsection{Using {\tt IM3SHAPE}}
\label{subsec:im3shape}
Using {\tt IM3SHAPE}, we measure the observed ellipticities of the galaxies in each field and use the ellipticity measurements to determine the galaxy position angles. When fitting a model of the galaxy shape to the galaxy image, {\tt IM3SHAPE} takes into account the contribution of the PSF by convolving the model galaxy with a star field image. The GREAT3 challenge provides star field images at the same resolution as the galaxy images for each field of the {\tt cgc} branch. When using {\tt IM3SHAPE} to measure the position angles of the galaxies, we use these star field images directly to calibrate for the PSF. However, when we require full ellipticity information (i.e. when estimating $f\left(\left|\bm{\epsilon}^{\mathrm{int}}\right|\right)$ in Section \ref{sec:g3_sims} and when using {\tt IM3SHAPE} to estimate the shear, as a comparison with the angle-only method, in Section \ref{sec:results}), we use star field images which are upsampled by a factor of seven using {\tt GalSim} \citep{rowe14}. We use original star field images for the position angle measurements to reduce computation time when calibrating the angle-only shear estimates using simulations. Since {\tt IM3SHAPE} corrects for the PSF when measuring the ellipticities, the angle-only method using {\tt IM3SHAPE} differs from the integrated light and moments based methods which do not include any PSF correction at this stage. Upon recovering measurements of $\bm{\epsilon}^{\mathrm{obs}}$, we estimate the position angles of the galaxies as
\begin{equation}\label{eq:im3shape_angles}
\hat{\alpha}=\frac{1}{2}\tan^{-1}\left(\frac{\epsilon_2^{\mathrm{obs}}}{\epsilon_1^{\mathrm{obs}}}\right).
\end{equation}

\section{Application to the GREAT3 simulations}
\label{sec:g3_sims}
The {\tt cgc} branch of the GREAT3 challenge simulates ground-based observations of 200 $10\times10\,\text{deg}^2$ fields. Each simulated observation contains $10^4$ resolved galaxy images. A constant shear is applied to all of the galaxies within a particular field. The images of the galaxies within each field are convolved with a constant anisotropic PSF, and a constant level of background Gaussian noise is assumed. The applied PSF and background noise levels are varied between the different observations, and the applied shear is different for each field observed. Noiseless star field images are included for each field providing an image of the PSF. The {\tt cgc} branch also includes five deep field observations for use as a training dataset. These observations consist of galaxy images that are one magnitude deeper than the challenge observations, but the dataset retains only images of the galaxies which would be present in the rest of the challenge. The properties of the simulations are discussed in detail in \cite{mandelbaum14}. 

In this section, we describe the application of the angle-only method to the {\tt cgc} branch of the GREAT3 simulations. We measure the position angles of the galaxies from the simulated images using the three methods discussed in the previous section. We begin by outlining the steps of the procedure followed in the angle-only analysis.

\begin{enumerate}
\item Calculate the $F_1\left(\left|\bm{g}\right|\right)$ function using equation (\ref{eq:general_F}). This requires an estimate of $f\left(\left|\bm{\epsilon}^{\mathrm{int}}\right|\right)$ which we obtain from the GREAT3 deep field calibration sets using {\tt IM3SHAPE}.
\item Measure the fluxes of the galaxies in the deep field images by summing over the pixel values and recover measurements of the half-light radii and the bulge to total flux ratio of the galaxies from {\tt IM3SHAPE}, to be used in calibration simulations.
\item Determine the relationship between the mean fluxes and the mean half-light radii of the five deep fields. 
\item Measure the mean flux of the galaxies in the field of interest and use this measurement to modify the fluxes and half-light radii of the deep field measurements output from step (ii) above.
\item Use the modified fluxes and half-light radii with the intrinsic ellipticity estimates, bulge to total flux ratio measurements, and a suite of random uniform distributed intrinsic position angles to simulate zero shear galaxy images using {\tt GalSim}. The noise in the simulated images is estimated from the GREAT3 image being analysed.
\item Measure the position angles of the galaxies in the zero shear simulations using one of the three methods discussed in the previous section. These measurements are used to construct a weighting function the purpose of which is to correct for PSF anisotropy and pixelization effects.
\item Measure the position angles of the galaxies in the field of interest and estimate the shear using weighted averages of the trigonometric functions.
\item Use the shear estimates to produce an updated set of simulations which include information about the shear.
\item Estimate the shear in the updated simulations using the same weighting function as in step (vi). Use these estimates to determine the bias in the estimates recovered in step (vii) and correct for the bias.
\item Repeat steps (vii)-(ix) using the corrected shear estimates as the input shear for the simulations until the estimated bias is below a desired threshold value.
\end{enumerate}

We now discuss our application of this procedure to the {\tt cgc} branch in detail.\\

\begin{figure*}
\begin{minipage}{6in}
\centering
\includegraphics{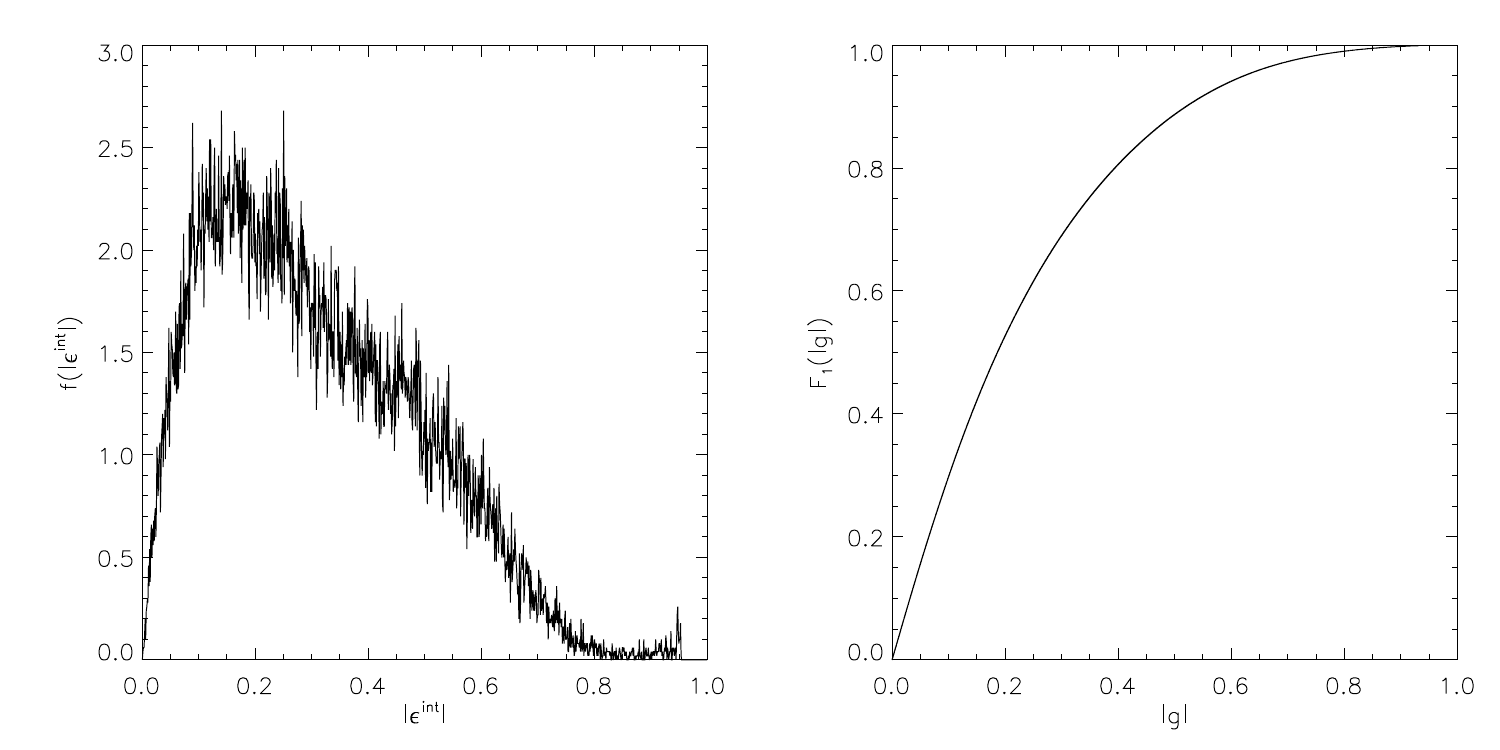}
\caption{Left-hand panel: The reconstructed $f\left(\left|\bm{\epsilon}^{\mathrm{int}}\right|\right)$ using {\tt IM3SHAPE} to recover the estimates of the intrinsic ellipticities of the galaxies in the deep field images of the {\tt cgc} branch and with a binwidth of $\Delta\left|\bm{\epsilon}^{\mathrm{int}}\right|=5\times10^{-3}$. Right-hand panel: The $F_1\left(\left|\bm{g}\right|\right)$ function calculated using equation (\ref{eq:general_F}) and the reconstructed $f\left(\left|\bm{\epsilon}^{\mathrm{int}}\right|\right)$ shown in the left-hand panel.}
\label{fig:pe_ffunc}
\end{minipage}
\end{figure*}

We began by estimating the form of the $F_1\left(\left|\bm{g}\right|\right)$ function (step (i)) using {\tt IM3SHAPE} to measure the observed ellipticities of the galaxies for each of the deep fields of the {\tt cgc} branch. When measuring a galaxy's ellipticity using {\tt IM3SHAPE}, we used star images which were upsampled by a factor of seven to correct for the PSF contribution (as opposed to the original star images used when we were concerned only with a galaxy's position angle). For each of the deep fields, we averaged over the observed ellipticity measurements to estimate the shear for that field. Using the estimated shears and the measured ellipticities, we recovered estimates of the intrinsic ellipticities of the galaxies by inverting equation (\ref{eq:observed_e}) and reconstructed $f\left(\left|\bm{\epsilon}^{\mathrm{int}}\right|\right)$, shown in Figure \ref{fig:pe_ffunc}. From this estimate of the distribution, we calculated the $F_1\left(\left|\bm{g}\right|\right)$ function numerically using equation (\ref{eq:general_F}); this is shown in the right panel of Figure \ref{fig:pe_ffunc}. As discussed in W14, the integral carried out when calculating the $F_1\left(\left|\bm{g}\right|\right)$ function smooths the $f\left(\left|\bm{\epsilon}^{\mathrm{int}}\right|\right)$ distribution. However, if we use a large binsize for the distribution we lose information. We therefore chose a binsize of $5\times10^{-3}$ in accordance with the binsize used in W14.

\begin{figure}
\centering
\includegraphics{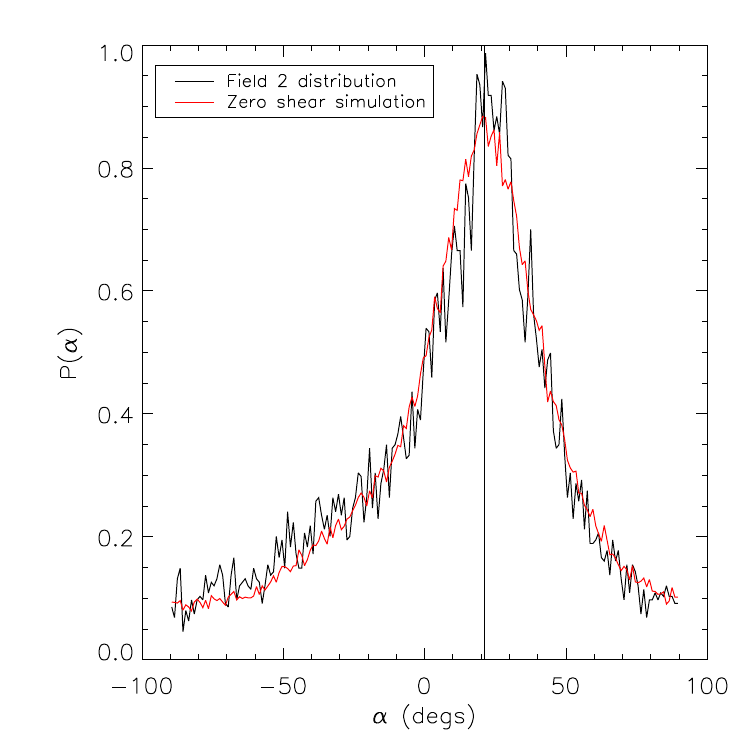}
\caption{The distribution of position angles measured from field 2 using the integrated light method (black curve). The vertical black line shows the position angle of the PSF. Here we see that the measured position angles are biased in the direction of the PSF. The red curve shows the distribution of measured position angles recovered from the zero shear simulations. We see that the distribution from the simulations provides a good description of the distribution from the image data, with the difference between the two being attributed to the shear signal for the zeroth-order shear estimate.}
\label{fig:field2_dist}
\end{figure}
 The effects of the PSF on a galaxy's observed ellipticity and orientation depend on the underlying ellipticity of the galaxy. When using the angle-only estimator, we do not recover information about the ellipticities, and therefore a complete understanding of the effects of the PSF on an individual galaxy's orientation cannot be realised. Instead, we chose to examine the impact of the PSF on the ensemble of galaxies. In the absence of a PSF (and pixelization), we expect the distribution of position angle measurements to be approximately uniform, with any deviation from uniformity being attributed to the underlying shear signal. In Figure \ref{fig:field2_dist}, we show the distribution of measured position angles from field 2 of the GREAT3 {\tt cgc} simulation set measured using the integrated light method. We see the unsurprising result that the position angles are biased in the direction of the PSF. This results in non-zero correlations between the galaxy position angles and the errors on the measurements which biases the shear estimates via equation (\ref{eq:cov_c_s_relation}).

To correct for this effect, we adopt a weighting scheme in which we downweight the contribution to the mean trigonometric functions (see equations (\ref{eq:est_ao}) and (\ref{eq:value_F})) from galaxies which align with the PSF. To understand how the galaxies should be weighted, let us assume a large sample of galaxies with a zero shear signal. In the absence of a PSF, we expect a uniform distribution of measured position angles and therefore the mean trigonometric functions to be zero. The observed unit vector of the galaxy is
\begin{equation}\label{eq:n_vec}
\bm{n}\left(\hat{\alpha}\right)=\left(
\begin{array}{c}
\cos\left(2\hat{\alpha}\right)\\
\sin\left(2\hat{\alpha}\right)
\end{array}\right).
\end{equation}
In the presence of a PSF, the angle distribution becomes non-uniform, and the mean unit vector, $\left<\hat{\bm{n}}\right>$, will be 
\begin{equation}\label{eq:mean_trig_psf}
\left<\hat{\bm{n}}\right>=\int_{-\frac{\pi}{2}}^{\frac{\pi}{2}}\mathrm{d}\hat{\alpha}\,\bm{n}\left(\hat{\alpha}\right)f_{\mathrm{PSF}}\left(\hat{\alpha}\right),
\end{equation}
where $f_{\mathrm{PSF}}\left(\hat{\alpha}\right)$ is the distribution of measured position angles given a non-zero PSF. If we introduce a weighting function, $w\left(\hat{\alpha}\right)$, such that
\begin{equation}\label{eq:mean_trig_psf_weight}
\left<\hat{\bm{n}}\right>=\int_{-\frac{\pi}{2}}^{\frac{\pi}{2}}\mathrm{d}\hat{\alpha}\,w\left(\hat{\alpha}\right)\bm{n}\left(\hat{\alpha}\right)f_{\mathrm{PSF}}\left(\hat{\alpha}\right),
\end{equation}
it is clear that one can correct for the effects of the PSF, such that we recover mean trigonometric functions equal to zero, if we set $w\left(\hat{\alpha}\right)=1/f_{\mathrm{PSF}}\left(\hat{\alpha}\right)$. To proceed, we therefore require an estimate of the distribution of measured position angles when a PSF is included.

To achieve this, we used {\tt GalSim} to simulate $10^5$ galaxy images with specific properties provided by {\tt IM3SHAPE} and assuming a zero input shear signal. For each galaxy in the five deep fields, we measured $\bm{\epsilon}^{\mathrm{obs}}$, the bulge to total flux ratio ($B/S$), and the half-light radius ($R_{\mathrm{e}}$). For each field, we also estimated the shear by averaging over $\bm{\epsilon}^{\mathrm{obs}}$ allowing us to estimate the $\bm{\epsilon}^{\mathrm{int}}$ of each galaxy by inverting equation (\ref{eq:observed_e}). For each of these galaxies, we simulate a corresponding galaxy using the measured properties as a template. The intensity profile of each simulated galaxy was assumed to be a S\'ersic profile consisting of a bulge with S\'ersic index $n_{\mathrm{s}}=4$ and a disc with $n_{\mathrm{s}}=1$, and with the half-light radii of each of these components being identical. The simulated galaxy is created with an intrinsic ellipticity of $\left|\bm{\epsilon}^{\mathrm{int}}\right|$ and with a position angle drawn randomly from a uniform distribution with the range $[-90^{\circ},90^{\circ})$. 

\begin{figure*}
\begin{minipage}{6in}
\centering
\includegraphics{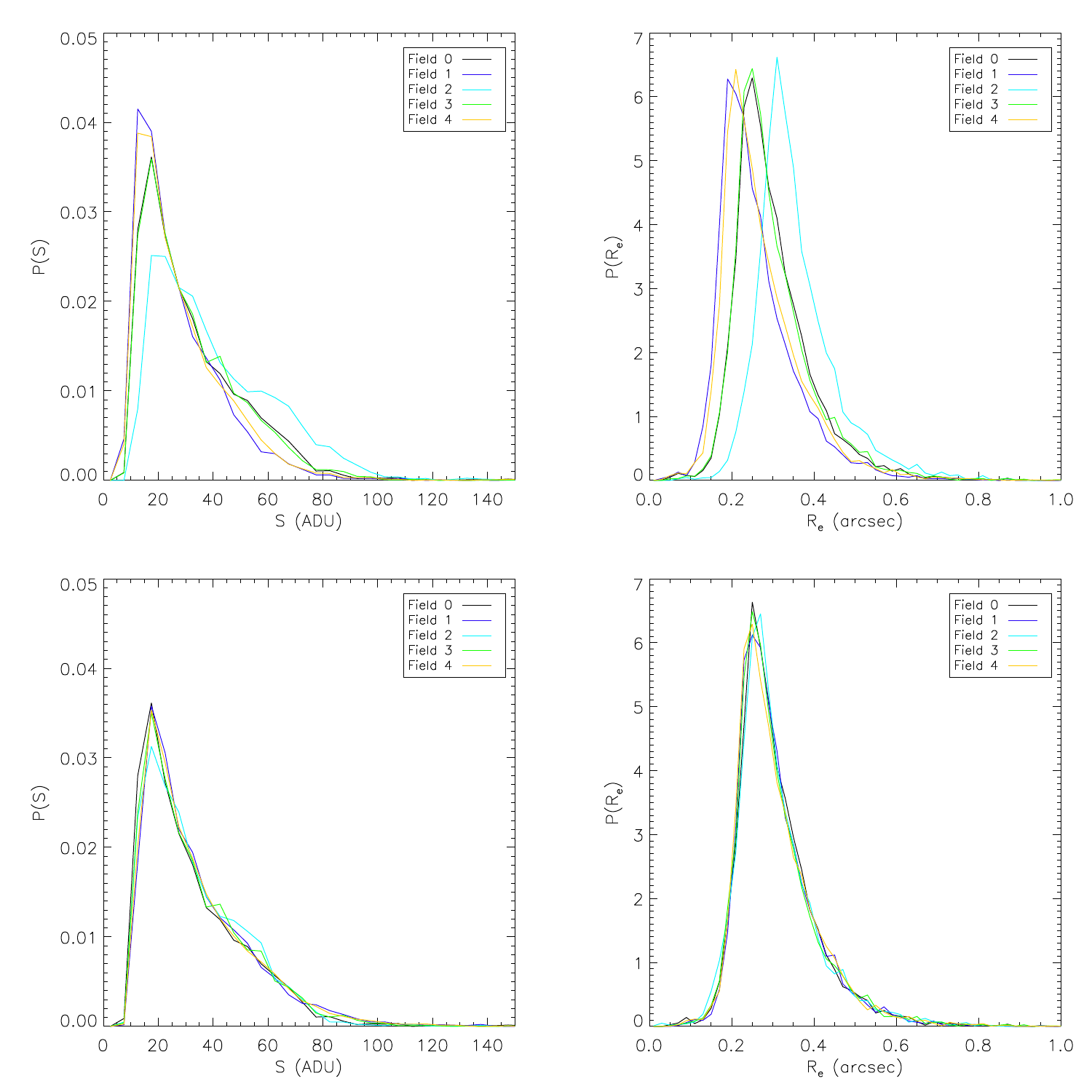}
\caption{The top left-hand panel shows the distributions of galaxy fluxes for the five deep fields of the {\tt cgc} branch. The top right-hand panel shows the distributions of half-light radii for the galaxies in these fields, with the half-light radii measured using {\tt IM3SHAPE}. In the bottom two panels we have modified the distributions of fields 1-4 using the relationships given by equations (\ref{eq:flux_relate}-\ref{eq:K_Re_relate}) so that they match the distributions of field 0. These plots indicate that the flux and half-light radii distributions in a particular field can be accurately reproduced using these relationships.}
\label{fig:deep_field_dists}
\end{minipage}
\end{figure*}
\begin{figure}
\centering
\includegraphics{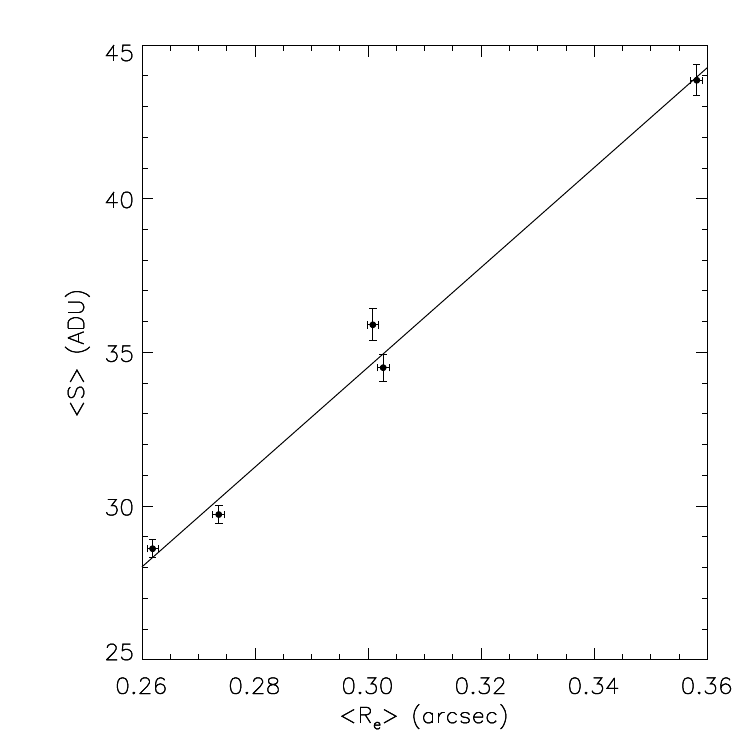}
\caption{The mean flux of the galaxies plotted as a function of the mean half-light radius for each of the five deep fields. Here we see a linear relationship between the two quantities. The line is the best fit to the data.}
\label{fig:mean_F_Re}
\end{figure}

For each field in the {\tt cgc} branch, the galaxy images are convolved with a different PSF, and only resolved galaxy images are included in the sample. Including only resolved galaxy images implies that the distributions of the half-light radii and, therefore, the fluxes of the galaxies are different for each field. In order to construct accurate calibration simulations, we require the distributions of these properties in the simulations to match the distributions in the field being analyzed. We used the properties estimated by {\tt IM3SHAPE} when analysing the deep field images to investigate how this can be achieved. First, we measured the flux of each galaxy by summing over the pixel values in each galaxy image, and we recovered the half-light radii estimates provided by {\tt IM3SHAPE} (step (ii)). Figure \ref{fig:deep_field_dists} shows the distributions of galaxy fluxes and half-light radii for each of the five deep fields. We see the difference between the distributions for each field. In Figure \ref{fig:mean_F_Re}, we plot the mean galaxy flux for each field as a function of the mean half-light radius. This indicates a linear relationship between the two quantities. We fitted a linear function of the form $\left<S\right>=m\left<R_{\mathrm{e}}\right>+c$ to this data (step (iii)) with the fitted parameters given by $m=162.4\,\mathrm{ADU}\,\mathrm{arcsec}^{-1}$ and $c=-14.19\,\mathrm{ADU}$. This relationship allows for a direct estimate of $\left<R_{\mathrm{e}}\right>$ simply by summing over the pixel values in a particular field, such that
\begin{equation}\label{eq:mean_hlr}
\left<R_{\mathrm{e}}\right>=\frac{\left<S\right>-c}{m}.
\end{equation}
One should not assign significance to the relationship shown in Figure \ref{fig:mean_F_Re}. The method for recovering estimates of $\left<R_{\mathrm{e}}\right>$ works as long as one can characterize a relation between $\left<R_{\mathrm{e}}\right>$ and $\left<S\right>$.

If $P_S^n\left(S\right)$ is the distribution of the fluxes in field $n$ and $\left<S\right>_n$ is the mean flux of a galaxy for that field, we find that we can write an approximate relationship for the flux distributions of two fields $a$ and $b$ as
\begin{equation}\label{eq:flux_relate}
P_S^a(S)=P_S^b(KS),
\end{equation}
where
\begin{equation}\label{eq:K_S_relate}
K=\frac{\left<S\right>_a}{\left<S\right>_b}.
\end{equation}
We find we can also write a similar relationship for the distributions of half-light radii of the two fields, $P_{R_{\mathrm{e}}}^a\left(R_{\mathrm{e}}\right)$ and $P_{R_{\mathrm{e}}}^b\left(R_{\mathrm{e}}\right)$,
\begin{equation}\label{eq:Re_relate}
P_{R_{\mathrm{e}}}^a(R_{\mathrm{e}})=P_{R_{\mathrm{e}}}^b(R_{\mathrm{e}}+K'),
\end{equation}
where
\begin{equation}\label{eq:K_Re_relate}
K'=\left<R_{\mathrm{e}}\right>_a-\left<R_{\mathrm{e}}\right>_b.
\end{equation}
If one measures the means of the fluxes of the galaxies in fields $a$ and $b$, one can recover estimates of the means of the half-light radii in these fields using the linear relationship shown in Figure \ref{fig:mean_F_Re} and equation (\ref{eq:mean_hlr}). It is then possible to modify the distributions of the fluxes and half-light radii in field $b$ so that they approximately match those in field $a$. An example of this procedure is shown in the bottom two panels of Figure \ref{fig:deep_field_dists}, where we have modified the distributions of the deep fields 1-4 to match deep field 0. We see that the distributions of the fluxes and half-light radii of a particular field can be accurately reproduced if the mean of the galaxy fluxes can be accurately measured and, as this requires simply summing over the image pixels within a particular field, we expect this to be achievable.

When applying this method to the challenge data, we measured the mean flux of the galaxies in the field of interest. We then modified the fluxes and half-light radii of all the galaxies in the five deep fields. We combined the modified fluxes and half-light radii with the estimated values of $\left|\bm{\epsilon}^{\mathrm{int}}\right|$ and $B/S$, and a suite of uniform random intrinsic position angles to provide us with the inputs required for $5\times10^4$ simulated galaxies (step (iv)).  The GREAT3 simulations use galaxy pairs whereby for each galaxy with an intrinsic orientation of $\alpha^{\mathrm{int}}$ there is an identical galaxy with an intrinsic orientation of $\alpha^{\mathrm{int}}+90^{\circ}$. This is done to reduce intrinsic shape noise in the shear estimates and hence the number of galaxies needed to average over. For each of our $5\times10^4$ simulated galaxies, we therefore created an identical galaxy image but rotated by $90^{\circ}$ with respect to the first, mimicking the procedure implemented by GREAT3. This provided us with $10^5$ simulated galaxy images. These images were convolved with the upsampled star field images to simulate the effects of the PSF. The pixel scale of the simulations was chosen to match the pixel scale used by GREAT3. The noise in the images was assumed to be Gaussian with the variance estimated from the values of the outermost pixels of each $48\times48\,\text{pixel}^2$ stamp in the field and with all $10^4$ stamps used in the estimate (step(v)).

Once the simulated galaxy images were created, we measured the position angles of the galaxies using each of the three methods discussed above (step (vi)). The results of this for field 2 when using the integrated light method are presented as the red curve in Figure \ref{fig:field2_dist}. This distribution was binned with a bin size of $1^{\circ}$ to give the distribution of position angles, $P\left(\alpha_i\right)$. As described above, the required weighting function is the reciprocal of this distribution
\begin{equation}\label{eq:discrete_dist}
w\left(\alpha_i\right)=\frac{1}{P\left(\alpha_i\right)}.
\end{equation}
We used this weighting function to correct the averages of the observed trigonometric functions for the PSF and pixelization, such that
\begin{equation}\label{eq:weighted_averages}
\left<\hat{\bm{n}}\right>=\frac{\sum_{i=1}^Nw\left(\hat{\alpha}_i\right)\hat{\bm{n}}_i}{\sum_{i=1}^Nw\left(\hat{\alpha}_i\right)}.
\end{equation}
Ideally, the weighting applied to each galaxy in the image would depend on the position angle of the galaxy without the effect of lensing (but with the effects of PSF, pixelization and noise still present). However, this is obviously not possible for real data. Assuming that the rotation induced by lensing is small, the weighting of the trigonometric functions will not be significantly affected by lensing and, provided the simulations are accurate, any resulting bias to the shear estimates should be corrected for by the iterative procedure discussed below.

Using the weighted averages given by equation (\ref{eq:weighted_averages}), we recover a zeroth-order estimate of the input shear signal, $\hat{\bm{g}}^{(0)}$, in each field via equations (\ref{eq:est_ao}) and (\ref{eq:value_F}) (step (vii)). Let $m_i'$ and $c_i'$  respectively denote the multiplicative and additive biases on the components of the zeroth-order shear estimates (\citealt{heymans06, huterer06, massey07}). The estimates can then be written as
\begin{equation}\label{eq:zeroth_bias}
\hat{g}_i^{(0)}=g_i+m_i'g_i+c_i'+\delta g_i^{(0)},
\end{equation}
where $\delta g_i^{(0)}$ is an error on the estimate with zero mean. If we assume that the weighting function successfully mitigates the effects of an additive bias, the zeroth-order estimate is expected to underestimate the modulus of the true shear signal as there is no correction for the measurement error bias arising from $\beta_c$ in equation (\ref{eq:cov_c_s}). It is possible to correct for this bias using the zero shear simulations to model the measurement error for each field as discussed in W14. However, for cases where there are large contributions from anisotropic PSFs, we find that the errors on the position angle measurements can be large leading to small values of $\beta_c$. Hence, attempting to correct for these biases can lead to substantial outliers in the shear estimates. Also, errors on the estimates of $\beta_c$ propagate nonlinearly into estimates of the shear. We, therefore, choose to simulate a further suite of $10^5$ galaxy images using the procedure described above, and we shear the galaxies using the estimates $\hat{\bm{g}}^{(0)}$ (step (viii)) for each field. We measure the shear from these updated simulations and use these estimates to determine, and correct for, the bias in the initial estimates recovered for each field (step (ix)), such that
\begin{equation}\label{eq:first_it}
\hat{\bm{g}}^{(1)}=\hat{\bm{g}}^{(0)}-\left(\hat{\bm{g}}_{\mathrm{sim}}^{(1)}-\hat{\bm{g}}^{(0)}\right),
\end{equation}
where the shear estimate recovered from the first-order simulations is
\begin{equation}\label{eq:first_it_clarify}
\hat{g}_{i,\mathrm{sim}}^{(1)}=\hat{g}_i^{(0)}+m_i'\hat{g}_i^{(0)}+c_i'+\delta g_i^{(1)}.
\end{equation}
This method, therefore, corrects for the bias introduced to the estimates of the shear and does not correct for the bias on the individual galaxy position angle estimates. We can iterate this step until estimates of the shear between subsequent iterations are consistent (step (x)).

For a large number of simulated galaxy images, noise in the estimated bias will be sub-dominant to noise in the zeroth-order shear estimate. However, using this iterative method, noise in the simulated shear estimates, $\hat{\bm{g}}_{\mathrm{sim}}^{(1)}$, due to a finite number of simulated galaxy images, propagates linearly into the final shear estimates. Therefore, if we assume that the details of the simulations are accurate, there is no additional noise bias expected from this procedure. For the $n^{\mathrm{th}}$ iteration, we can write
\begin{equation}\label{eq:nth_it}
\hat{\bm{g}}^{(n)}=\hat{\bm{g}}^{(0)}-\left(\hat{\bm{g}}_{\mathrm{sim}}^{(n)}-\hat{\bm{g}}^{(n-1)}\right).
\end{equation}
Assuming that the simulations provide an accurate description of the true field, we show in Appendix A that the residual bias for the $n^{\mathrm{th}}$ iteration is
\begin{equation}\label{eq:bias_it}
\left<\hat{g}_i^{(n)}-g_i\right>=\left(-m_i'\right)^{n}\left(m_i'g_i+c_i')\right),
\end{equation}
which converges to zero for $\left|m_i'\right|<1$. As explained above, the zeroth-order shear estimates are expected to underestimate the modulus of the true shear. Hence, we expect $m_i'$ to be confined to the range $-1<m_i'<0$, and therefore the shear estimates to converge for all fields.

\section{Results}
\label{sec:results}
Here we compare the results of the angle-only approach using each of the three methods to measure the position angles of the galaxies discussed in Section \ref{sec:ang_meas}. We include the results obtained from a naive application of {\tt IM3SHAPE} where full ellipticity information is used (measured using the upsampled star images to correct for the PSF) with no additional calibration scheme. 

The $Q$-value is the metric used to quantify the performance of the estimators in the GREAT3 challenge \citep{mandelbaum14}. It is defined as a function of the multiplicative and additive biases ($m$ and $c$ respectively) where the biases satisfy the approximation (\citealt{heymans06, huterer06, massey07})
\begin{equation}\label{eq:m_c}
\hat{g}_i=\left(1+m_i\right)g_i+c_i+\delta g_i.
\end{equation}
The subscript $i$ denotes the components of the shear in a reference frame aligned with the PSF in the field being analysed. The biases are estimated using a linear regression over the 200 fields given the known input shears. The $Q$ value is then calculated as 
\begin{equation}
Q=\frac{2000\eta}{\sqrt{\sigma_{\text{min}}^2+\sum_{i={+,\times}}\left(\frac{m_i}{m_{\text{target}}}\right)^2+\sum_{i={+,\times}}\left(\frac{c_i}{c_{\text{target}}}\right)^2}},
\end{equation}
where the subscript $+$ corresponds to the components of the biases in the direction aligned with the PSF for each field and the subscript $\times$ denotes the direction perpendicular to the PSF. The target values of the biases, $m_{\text{target}}$ and $c_{\text{target}}$, are based on the requirements of the ESA Euclid space mission \citep{massey13} and are given as $m_{\text{target}}=2\times10^{-3}$ and $c_{\text{target}}=2\times10^{-4}$. The constant $\eta$ normalizes the metric such that a value of $Q\approx1000$ is expected for estimates of the shear which achieve the target values of $m$ and $c$. The term $\sigma_{\text{min}}^2$ corresponds to the typical dispersion of the biases due to pixel noise and is determined using trial submissions to the GREAT3 challenge. For the challenge, the values adopted for the {\tt cgc} branch were $\eta=1.232$ and $\sigma_{\text{min}}^2=4$. We estimated the $Q$-values for our methods using the publicly available GREAT3 metric evaluation script\footnote{\url{https://github.com/barnabytprowe/great3-public/wiki/Metric-evaluation-scripts-and-truth-data}}. This script also provides the estimates of $m_i$ and $c_i$ used to calculate the $Q$-value and their corresponding error bars.

In the discussion that follows, we include the highest Q-value entry to the {\tt cgc} branch of the GREAT3 challenge\footnote{\url{http://great3.jb.man.ac.uk/leaderboard/board/post\_challenge/24}} from {\tt IM3SHAPE} - for a comparison of the performance of the angle-only method with the challenge submission of a model based method. This entry implements a multiplicative calibration factor to correct for noise biases expected to arise in Maximum-Likelihood shape estimation. Additive biases are expected in the presence of anisotropic PSFs. However, no calibration for this bias is included. We also include the highest entry submitted to the challenge using the KSB method - for a comparison with a moments based method. The details of the submissions using {\tt IM3SHAPE} and the KSB method are presented in \citep{mandelbaum14b}. For the angle-only analyses, the results presented made use of a single iteration of the procedure discussed in Section \ref{sec:g3_sims}.

\begin{figure*}
\begin{minipage}{6in}
\centering
\includegraphics{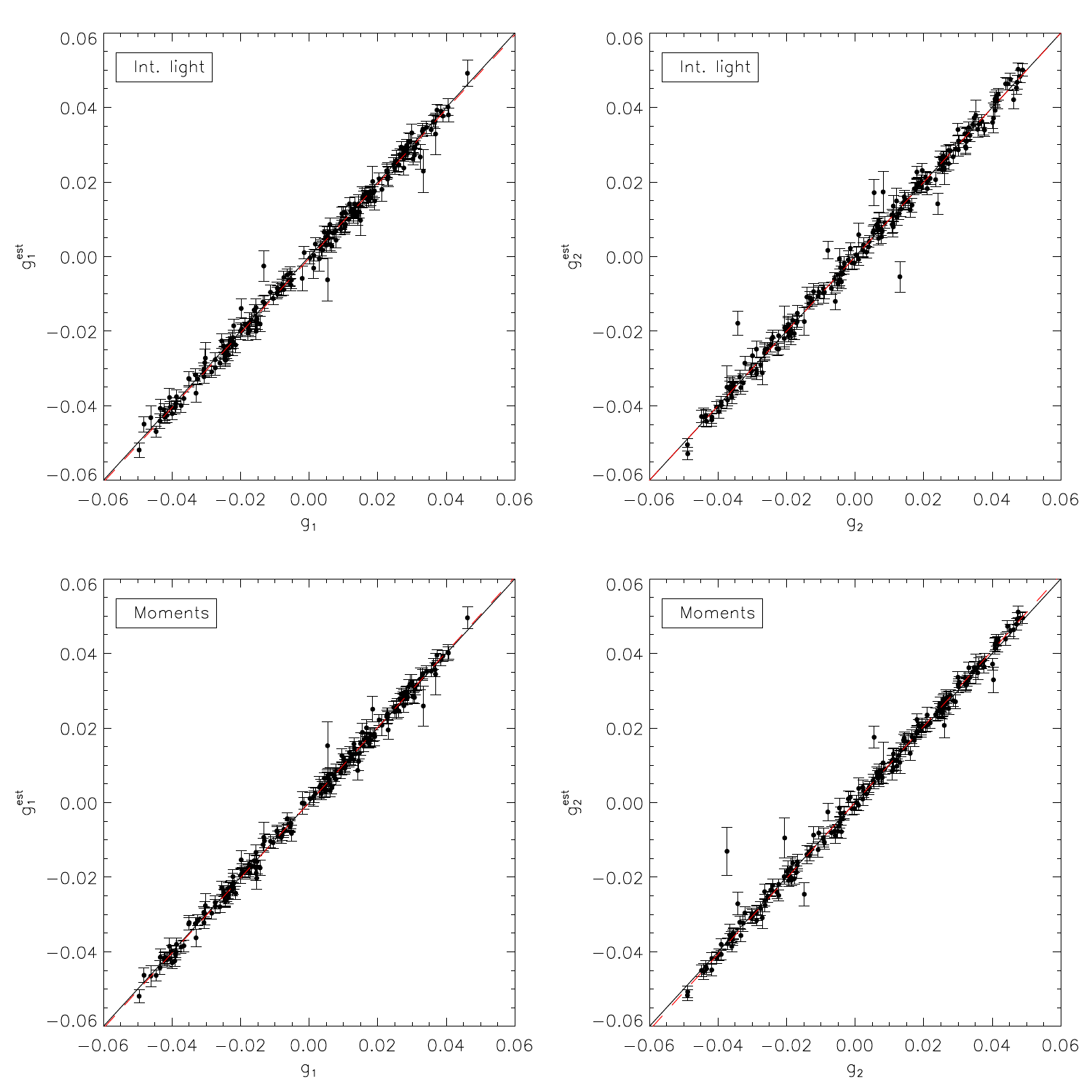}
\caption{The top panels show the recovered shear estimates for each of the 200 fields of the {\tt cgc} branch plotted against the input shear values and using the integrated light method to measure the position angles. The bottom panels show the recovered shear estimates using the moments based method to measure the position angles. In all cases, the black line is the one-to-one line and the red-dashed line is the line of best fit for the recovered shear values.}
\label{fig:plot_int_mom}
\end{minipage}
\end{figure*}
\begin{figure*}
\begin{minipage}{6in}
\centering
\includegraphics{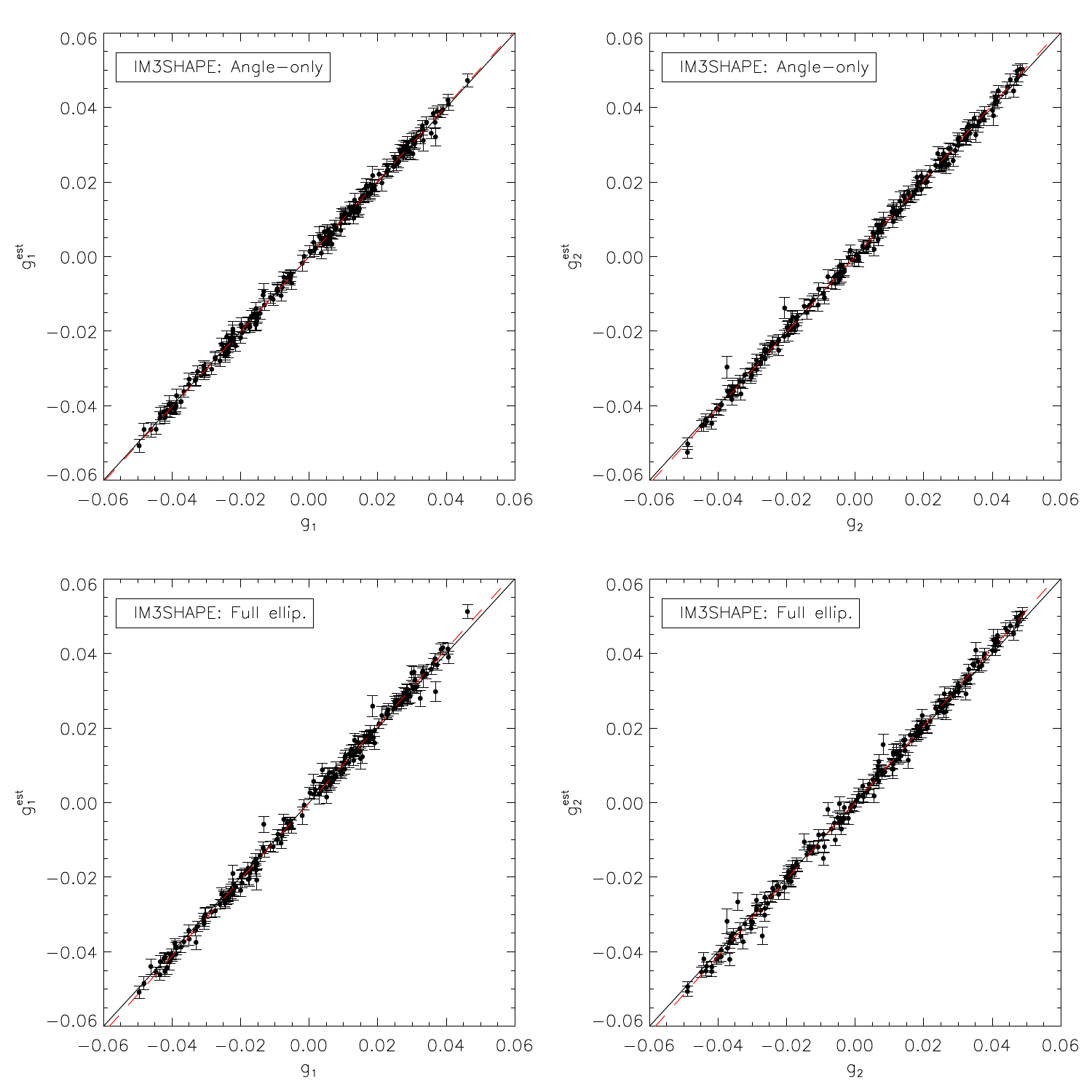}
\caption{The top panels show the shear estimates recovered using the angle-only method for each of the 200 fields of the {\tt cgc} branch plotted against the input shear values and using {\tt IM3SHAPE} to measure the position angles of the galaxies. The bottom panels show the recovered shear estimates using the full ellipticity information from {\tt IM3SHAPE} but with no additional calibration scheme. In all cases, the black line is the one-to-one line and the red-dashed line is the line of best fit for the recovered shear values.}
\label{fig:plot_im3}
\end{minipage}
\end{figure*}

Figure \ref{fig:plot_int_mom} shows the recovered angle-only shear estimates when using the integrated light and moments based methods to measure the position angles. These plots display the estimates for the components of the shear in a reference frame aligned with the simulated images. It should be emphasized that this is not the reference frame used to measure the $Q$-value, which employs a coordinate system that is aligned with the PSF for that field as described above. The error bars have been estimated using a linear approximation of the $F_1\left(\left|\bm{g}\right|\right)$ function as discussed in Appendix B. Figure \ref{fig:plot_im3} shows the results of {\tt IM3SHAPE} using both the angle-only method and full ellipticity information - where no further calibration scheme is implemented. We see that the errors are reduced for the angle-only method when using the ellipticities from {\tt IM3SHAPE} to measure the position angles as compared with the integrated light and moments based methods. This reduction is due to the correction for the PSF when measuring the ellipticities which increases the uniformity of the weighting function and reduces the errors on the angle measurements. Hence, the value of $\sum_{i=1}^Nw_i^2$ is lower, and the multiplicative bias for the zeroth-order estimate is smaller (or equivalently, the $\beta_{\mathrm{wc}}$ term as defined in Appendix B is larger) reducing the errors on the shear estimates in accordance with equations (\ref{eq:err_pairs}) and (\ref{eq:err_est_it}).

The absence of the PSF correction at the measurement stage also leads to a larger bias in the zeroth-order shear estimates when using the integrated light and the moments based methods. Therefore, a larger number of iterations are required to remove the bias in these methods as shown in equation (\ref{eq:bias_it}). The scatter on the shear estimates recovered using full ellipticity information is smaller than when using only the position angles. This is partly due to form of the $F_1\left(\left|\bm{g}\right|\right)$ function, as discussed in W14. However, there is also an increase in the errors due to the iterative procedure employed to remove the biases in the angle-only method; this is quantified in equation (\ref{eq:err_est_it}).

For all methods using the angle-only estimator, a residual bias is expected when using a finite number of iterations (for these tests we use a single iteration). However, the limiting factor in this method is expected to be the accuracy with which we can simulate the galaxies in the branch.

\begin{table*}
\begin{minipage}{6in}
\centering
\begin{tabular}{|c|c|c|c|c|c|}
\hline
Method & $Q$ & $m_+\,(\times10^{-3})$ & $m_{\times}\,(\times10^{-3})$ & $c_+\,(\times10^{-4})$ & $c_{\times}\,(\times10^{-4})$ \\ [1ex]
\hline
Int. light & $324^{+115}_{-110}$ & $-11.26\pm8.25$ & $-7.34\pm6.22$ & $-4.03\pm2.08$ & $4.34\pm1.58$ \\ [1ex]
Moments & $403^{+104}_{-101}$ & $2.44\pm7.96$ & $9.38\pm4.83$ & $5.32\pm2.01$ & $3.32\pm1.23$ \\ [1ex]
{\tt IM3SHAPE}: Angle-only & $371^{+96}_{-98}$ & $6.80\pm3.88$ & $10.35\pm4.05$ & $1.89\pm0.98$ & $1.78\pm1.03$ \\ [1ex] 
{\tt IM3SHAPE}: Full ellip. & $117_{-13}^{+13}$ & $29.47\pm5.18$ & $27.35\pm4.31$ & $-11.33\pm1.31$ & $0.36\pm1.10$ \\ [1ex]
{\tt IM3SHAPE}: Highest entry & $416_{-50}^{+49}$ & $0.00\pm5.00$ & $-2.09\pm4.14$ & $-11.0\pm1.26$ & $0.44\pm1.05$ \\[1ex]
KSB: Highest entry & $122_{-19}^{+18}$ & $22.7\pm7.3$ & $32.5\pm5.9$ & $6.19\pm1.85$ & $-1.07\pm1.51$ \\[1ex]
\hline
\end{tabular}
\caption{The results of the analyses we have performed on the GREAT3 {\tt cgc} branch. The first column shows the $Q$-values achieved for each method. The next two columns are the multiplicative biases estimated for each method. The final two columns show the estimated additive biases.}
\label{table:comp_ao_all_im3}
\end{minipage}
\end{table*}
\begin{figure*}
\begin{minipage}{6in}
\centering
\includegraphics{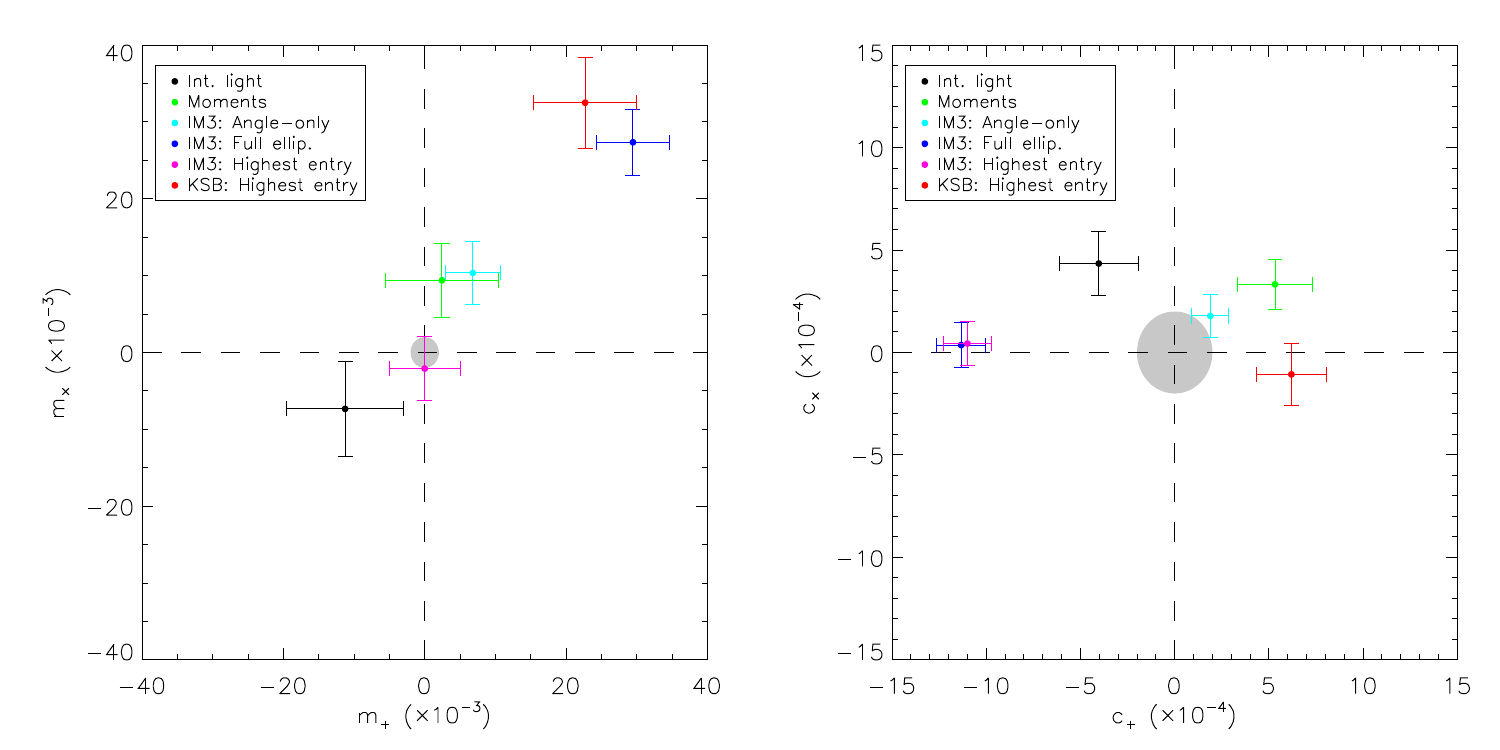}
\caption{The multiplicative and additive biases for each test on the {\tt cgc} branch. The grey region indicates the target bias values for future precision cosmic shear experiments, which are based on the target values evaluated for the Euclid space mission \citep{massey13}.}
\label{fig:plot_ms_cs}
\end{minipage}
\end{figure*}

The results of our analyses of the {\tt cgc} branch are presented in Table \ref{table:comp_ao_all_im3}. The error bars on each $Q$-value were determined by assuming that the estimates of $m_i$ and $c_i$ are the true values. We used simulations to confirm that the errors on the $m_i$ and $c_i$ estimates are approximately Gaussian. We then simulated a suite of $5\times10^5$ uncorrelated Gaussian random variables distributed about each of the estimated bias parameters with the dispersions provided by the metric evaluation script. These simulated bias parameters were used to calculate the $5\times10^5$ corresponding $Q$-values using the least squares method, and then a histogram of these values was constructed. The error bars were calculated as the $68.3\%$ confidence regions of the simulated $Q$-values distributed about the mean recovered value.

From Table \ref{table:comp_ao_all_im3} we see that the $Q$-values for the three angle-only analyses are competitive with the highest entry submitted using {\tt IM3SHAPE} to within one standard deviation. The $Q$-values achieved in all of the analyses performed with the angle-only approach are greater than the highest entry submitted using the KSB method. Figure \ref{fig:plot_ms_cs} displays the values of $m_i$ and $c_i$ calculated using the GREAT3 metric evaluation script. The plus subscript corresponds to the direction parallel with the PSF and the cross subscript to the direction perpendicular to the PSF. We see that the multiplicative biases of the integrated light and moments based angle-only methods lie within two standard deviations of the target values identified for future ``Stage IV" weak lensing experiments aimed at precision dark energy constraints. There appears to be a greater residual additive bias when using these methods to measure the position angles than when using the angles recovered by {\tt IM3SHAPE}. This is likely to be due to there being no correction for the PSF during the angle measurement stage, and hence there is a larger additive bias for the zeroth-order shear estimate. Assuming that this is the case, the bias should decrease as the number of iterations is increased, as indicated by equation (\ref{eq:bias_it}), until the threshold enforced by the accuracy of the simulations is reached. The angle-only method using {\tt IM3SHAPE} exhibits a reduction in the multiplicative bias as compared with that of the naive full ellipticity approach, and we also see a reduction in the component of the additive bias which aligns with the PSF. 
  
\section{Discussion}
\label{sec:discussion}
We have demonstrated an algorithm for applying the angle-only weak lensing estimator of W14 to realistic galaxy images. We find that the performance of this technique is competitive with state-of-the-art shape measurement techniques. To measure the position angles of the galaxies, we considered three separate approaches: the integrated light method, a moments based method, and using ellipticities measured by {\tt IM3SHAPE}. All three methods yield $Q$-values consistent with those of the highest {\tt IM3SHAPE} submission and greater than the highest KSB submission.

The angle-only estimator requires a measurement of the $F_1\left(\left|\bm{g}\right|\right)$ function. We have shown that this function can be successfully recovered from deep field images using {\tt IM3SHAPE}. Constraints on the accuracy of the recovered $F_1\left(\left|\bm{g}\right|\right)$ function required to provide reliable shear estimates are discussed in W14. When measuring the ellipticities of the galaxies in the deep field images using {\tt IM3SHAPE}, we necessarily fit for the half-light radii and the bulge to total flux ratios of the galaxies. We have introduced a method of using this information to construct accurate calibration simulations. To do this, we modify the distributions of the fluxes and half-light radii measured from the deep fields to match the distributions in the field being analysed by simply summing over the pixel values in that field. This approach could be useful for any analysis which requires calibration simulations. It is, as yet, unclear as to how general the calibration methods developed in this paper are with regards to how the galaxies of a particular survey are selected. A detailed investigation of this issue is left for future work. However, provided that one can construct accurate zero shear calibration simulations, the multiplicative and additive biases present in the zeroth-order shear estimates can be corrected for using the iterative approach.  

The presence of an anisotropic PSF and pixelization bias position angle estimates. We have employed a weighting scheme to reduce this effect. The weighting function is estimated using calibration simulations which assume a zero input shear signal. Using the iterative method employed for this paper, we have argued that residual biases in the shear estimates can be reduced below a threshold which is determined by the accuracy of the calibration simulations. For a perfect suite of calibration simulations, the multiplicative bias in the zeroth-order shear estimate is essentially due to noise in the image, pixelization and the PSF. The magnitude of the multiplicative bias in the zeroth-order shear estimate determines the rate of convergence for the estimator. Therefore, correcting for the PSF prior to the zeroth-order estimate should increase the rate of convergence. However, the effect of the PSF on a galaxy's orientation is model dependent and is, therefore, difficult to correct for when using an angle measurement method which is independent of the ellipticity. 

The purpose of this paper is to demonstrate the feasibility of performing an angle-only shear analysis on realistic galaxy images. To demonstrate this, we have focused on the simulated images of the {\tt cgc} branch of the GREAT3 challenge. In future work, we will look at how the angle-only method can be applied to fields with a variable shear and a variable PSF. We also aim to reduce the dependence of the method on simulations and develop a deeper understanding of the level at which the angle-only method can complement ellipticity based methods and ultimately reduce systematics in weak lensing surveys.

\section*{Acknowledgments}
We thank Joe Zuntz for useful comments. LW and MLB are grateful to the ERC for support through the award of an
ERC Starting Independent Researcher Grant (EC FP7 grant number 280127). MLB also thanks the STFC for the award of Advanced and
Halliday fellowships (grant number ST/I005129/1). 

\bibliographystyle{mn2e}
\bibliography{ms}

\begin{thebibliography}{20}
\expandafter\ifx\csname natexlab\endcsname\relax\def\natexlab#1{#1}\fi

\bibitem[{{Bartelmann} \& {Schneider}(2001)}]{bartelmann01}
{Bartelmann} M., {Schneider} P., 2001, \physrep, 340, 291

\bibitem[{{Brown} {et~al.}(2003){Brown}, {Taylor}, {Bacon}, {Gray}, {Dye},
  {Meisenheimer}, \& {Wolf}}]{brown03}
{Brown} M.~L., {Taylor} A.~N., {Bacon} D.~J., {Gray} M.~E., {Dye} S.,
  {Meisenheimer} K., {Wolf} C., 2003, \mnras, 341, 100

\bibitem[{{Fu} {et~al.}(2008){Fu}, {Semboloni}, {Hoekstra}, {Kilbinger}, {van
  Waerbeke}, {Tereno}, {Mellier}, {Heymans}, {Coupon}, {Benabed}, {Benjamin},
  {Bertin}, {Dor{\'e}}, {Hudson}, {Ilbert}, {Maoli}, {Marmo}, {McCracken}, \&
  {M{\'e}nard}}]{fu08}
{Fu} L., {Semboloni} E., {Hoekstra} H., {Kilbinger} M., {van Waerbeke} L.,
  {Tereno} I., {Mellier} Y., {Heymans} C., {Coupon} J., {Benabed} K.,
  {Benjamin} J., {Bertin} E., {Dor{\'e}} O., {Hudson} M.~J., {Ilbert} O.,
  {Maoli} R., {Marmo} C., {McCracken} H.~J., {M{\'e}nard} B., 2008, \aap, 479,
  9

\bibitem[{{Heymans} {et~al.}(2006){Heymans}, {Van Waerbeke}, {Bacon}, {Berge},
  {Bernstein}, {Bertin}, {Bridle}, {Brown}, {Clowe}, {Dahle}, {Erben}, {Gray},
  {Hetterscheidt}, {Hoekstra}, {Hudelot}, {Jarvis}, {Kuijken}, {Margoniner},
  {Massey}, {Mellier}, {Nakajima}, {Refregier}, {Rhodes}, {Schrabback}, \&
  {Wittman}}]{heymans06}
{Heymans} C., {Van Waerbeke} L., {Bacon} D., {Berge} J., {Bernstein} G.,
  {Bertin} E., {Bridle} S., {Brown} M.~L., {Clowe} D., {Dahle} H., {Erben} T.,
  {Gray} M., {Hetterscheidt} M., {Hoekstra} H., {Hudelot} P., {Jarvis} M.,
  {Kuijken} K., {Margoniner} V., {Massey} R., {Mellier} Y., {Nakajima} R.,
  {Refregier} A., {Rhodes} J., {Schrabback} T., {Wittman} D., 2006, \mnras,
  368, 1323

\bibitem[{{Hoekstra} {et~al.}(2006){Hoekstra}, {Mellier}, {van Waerbeke},
  {Semboloni}, {Fu}, {Hudson}, {Parker}, {Tereno}, \& {Benabed}}]{hoekstra06}
{Hoekstra} H., {Mellier} Y., {van Waerbeke} L., {Semboloni} E., {Fu} L.,
  {Hudson} M.~J., {Parker} L.~C., {Tereno} I., {Benabed} K., 2006, \apj, 647,
  116

\bibitem[{{Huterer} {et~al.}(2006){Huterer}, {Takada}, {Bernstein}, \&
  {Jain}}]{huterer06}
{Huterer} D., {Takada} M., {Bernstein} G., {Jain} B., 2006, \mnras, 366, 101

\bibitem[{{Kaiser} {et~al.}(1995){Kaiser}, {Squires}, \&
  {Broadhurst}}]{kaiser95}
{Kaiser} N., {Squires} G., {Broadhurst} T., 1995, \apj, 449, 460

\bibitem[{{Kilbinger} {et~al.}(2013){Kilbinger}, {Fu}, {Heymans}, {Simpson},
  {Benjamin}, {Erben}, {Harnois-D{\'e}raps}, {Hoekstra}, {Hildebrandt},
  {Kitching}, {Mellier}, {Miller}, {Van Waerbeke}, {Benabed}, {Bonnett},
  {Coupon}, {Hudson}, {Kuijken}, {Rowe}, {Schrabback}, {Semboloni}, {Vafaei},
  \& {Velander}}]{kilbinger13}
{Kilbinger} M., {Fu} L., {Heymans} C., {Simpson} F., {Benjamin} J., {Erben} T.,
  {Harnois-D{\'e}raps} J., {Hoekstra} H., {Hildebrandt} H., {Kitching} T.~D.,
  {Mellier} Y., {Miller} L., {Van Waerbeke} L., {Benabed} K., {Bonnett} C.,
  {Coupon} J., {Hudson} M.~J., {Kuijken} K., {Rowe} B., {Schrabback} T.,
  {Semboloni} E., {Vafaei} S., {Velander} M., 2013, \mnras, 430, 2200

\bibitem[{{Kitching} {et~al.}(2008){Kitching}, {Miller}, {Heymans}, {van
  Waerbeke}, \& {Heavens}}]{kitching08}
{Kitching} T.~D., {Miller} L., {Heymans} C.~E., {van Waerbeke} L., {Heavens}
  A.~F., 2008, \mnras, 390, 149

\bibitem[{{Mandelbaum} {et~al.}(2014{\natexlab{a}}){Mandelbaum}, {Rowe},
  {Armstrong}, {Bard}, {Bertin}, {Bosch}, {Boutigny}, {Courbin}, {Dawson},
  {Donnarumma}, {Fenech Conti}, {Gavazzi}, {Gentile}, {Gill}, {Hogg}, {Huff},
  {Jee}, {Kacprzak}, {Kilbinger}, {Kuntzer}, {Lang}, {Luo}, {March},
  {Marshall}, {Meyers}, {Miller}, {Miyatake}, {Nakajima}, {Ngole Mboula},
  {Nurbaeva}, {Okura}, {Paulin-Henriksson}, {Rhodes}, {Schneider}, {Shan},
  {Sheldon}, {Simet}, {Starck}, {Sureau}, {Tewes}, {Zarb Adami}, {Zhang}, \&
  {Zuntz}}]{mandelbaum14b}
{Mandelbaum} R., {Rowe} B., {Armstrong} R., {Bard} D., {Bertin} E., {Bosch} J.,
  {Boutigny} D., {Courbin} F., {Dawson} W.~A., {Donnarumma} A., {Fenech Conti}
  I., {Gavazzi} R., {Gentile} M., {Gill} M.~S.~S., {Hogg} D.~W., {Huff} E.~M.,
  {Jee} M.~J., {Kacprzak} T., {Kilbinger} M., {Kuntzer} T., {Lang} D., {Luo}
  W., {March} M.~C., {Marshall} P.~J., {Meyers} J.~E., {Miller} L., {Miyatake}
  H., {Nakajima} R., {Ngole Mboula} F.~M., {Nurbaeva} G., {Okura} Y.,
  {Paulin-Henriksson} S., {Rhodes} J., {Schneider} M.~D., {Shan} H., {Sheldon}
  E.~S., {Simet} M., {Starck} J.-L., {Sureau} F., {Tewes} M., {Zarb Adami} K.,
  {Zhang} J., {Zuntz} J., 2014{\natexlab{a}}, ArXiv e-prints

\bibitem[{{Mandelbaum} {et~al.}(2014{\natexlab{b}}){Mandelbaum}, {Rowe},
  {Bosch}, {Chang}, {Courbin}, {Gill}, {Jarvis}, {Kannawadi}, {Kacprzak},
  {Lackner}, {Leauthaud}, {Miyatake}, {Nakajima}, {Rhodes}, {Simet}, {Zuntz},
  {Armstrong}, {Bridle}, {Coupon}, {Dietrich}, {Gentile}, {Heymans}, {Jurling},
  {Kent}, {Kirkby}, {Margala}, {Massey}, {Melchior}, {Peterson}, {Roodman}, \&
  {Schrabback}}]{mandelbaum14}
{Mandelbaum} R., {Rowe} B., {Bosch} J., {Chang} C., {Courbin} F., {Gill} M.,
  {Jarvis} M., {Kannawadi} A., {Kacprzak} T., {Lackner} C., {Leauthaud} A.,
  {Miyatake} H., {Nakajima} R., {Rhodes} J., {Simet} M., {Zuntz} J.,
  {Armstrong} B., {Bridle} S., {Coupon} J., {Dietrich} J.~P., {Gentile} M.,
  {Heymans} C., {Jurling} A.~S., {Kent} S.~M., {Kirkby} D., {Margala} D.,
  {Massey} R., {Melchior} P., {Peterson} J., {Roodman} A., {Schrabback} T.,
  2014{\natexlab{b}}, \apjs, 212, 5

\bibitem[{{Massey} {et~al.}(2007){Massey}, {Heymans}, {Berg{\'e}}, {Bernstein},
  {Bridle}, {Clowe}, {Dahle}, {Ellis}, {Erben}, {Hetterscheidt}, {High},
  {Hirata}, {Hoekstra}, {Hudelot}, {Jarvis}, {Johnston}, {Kuijken},
  {Margoniner}, {Mandelbaum}, {Mellier}, {Nakajima}, {Paulin-Henriksson},
  {Peeples}, {Roat}, {Refregier}, {Rhodes}, {Schrabback}, {Schirmer}, {Seljak},
  {Semboloni}, \& {van Waerbeke}}]{massey07}
{Massey} R., {Heymans} C., {Berg{\'e}} J., {Bernstein} G., {Bridle} S., {Clowe}
  D., {Dahle} H., {Ellis} R., {Erben} T., {Hetterscheidt} M., {High} F.~W.,
  {Hirata} C., {Hoekstra} H., {Hudelot} P., {Jarvis} M., {Johnston} D.,
  {Kuijken} K., {Margoniner} V., {Mandelbaum} R., {Mellier} Y., {Nakajima} R.,
  {Paulin-Henriksson} S., {Peeples} M., {Roat} C., {Refregier} A., {Rhodes} J.,
  {Schrabback} T., {Schirmer} M., {Seljak} U., {Semboloni} E., {van Waerbeke}
  L., 2007, \mnras, 376, 13

\bibitem[{{Massey} {et~al.}(2013){Massey}, {Hoekstra}, {Kitching}, {Rhodes},
  {Cropper}, {Amiaux}, {Harvey}, {Mellier}, {Meneghetti}, {Miller},
  {Paulin-Henriksson}, {Pires}, {Scaramella}, \& {Schrabback}}]{massey13}
{Massey} R., {Hoekstra} H., {Kitching} T., {Rhodes} J., {Cropper} M., {Amiaux}
  J., {Harvey} D., {Mellier} Y., {Meneghetti} M., {Miller} L.,
  {Paulin-Henriksson} S., {Pires} S., {Scaramella} R., {Schrabback} T., 2013,
  \mnras, 429, 661

\bibitem[{{Melchior} {et~al.}(2011){Melchior}, {Viola}, {Sch{\"a}fer}, \&
  {Bartelmann}}]{melchior11}
{Melchior} P., {Viola} M., {Sch{\"a}fer} B.~M., {Bartelmann} M., 2011, \mnras,
  412, 1552

\bibitem[{{Miller} {et~al.}(2007){Miller}, {Kitching}, {Heymans}, {Heavens}, \&
  {van Waerbeke}}]{miller07}
{Miller} L., {Kitching} T.~D., {Heymans} C., {Heavens} A.~F., {van Waerbeke}
  L., 2007, \mnras, 382, 315

\bibitem[{{Rowe} {et~al.}(2014){Rowe}, {Jarvis}, {Mandelbaum}, {Bernstein},
  {Bosch}, {Simet}, {Meyers}, {Kacprzak}, {Nakajima}, {Zuntz}, {Miyatake},
  {Dietrich}, {Armstrong}, {Melchior}, \& {Gill}}]{rowe14}
{Rowe} B., {Jarvis} M., {Mandelbaum} R., {Bernstein} G.~M., {Bosch} J., {Simet}
  M., {Meyers} J.~E., {Kacprzak} T., {Nakajima} R., {Zuntz} J., {Miyatake} H.,
  {Dietrich} J.~P., {Armstrong} R., {Melchior} P., {Gill} M.~S.~S., 2014, ArXiv
  e-prints

\bibitem[{{Schrabback} {et~al.}(2010){Schrabback}, {Hartlap}, {Joachimi},
  {Kilbinger}, {Simon}, {Benabed}, {Brada{\v c}}, {Eifler}, {Erben},
  {Fassnacht}, {High}, {Hilbert}, {Hildebrandt}, {Hoekstra}, {Kuijken},
  {Marshall}, {Mellier}, {Morganson}, {Schneider}, {Semboloni}, {van Waerbeke},
  \& {Velander}}]{schrabback10}
{Schrabback} T., {Hartlap} J., {Joachimi} B., {Kilbinger} M., {Simon} P.,
  {Benabed} K., {Brada{\v c}} M., {Eifler} T., {Erben} T., {Fassnacht} C.~D.,
  {High} F.~W., {Hilbert} S., {Hildebrandt} H., {Hoekstra} H., {Kuijken} K.,
  {Marshall} P.~J., {Mellier} Y., {Morganson} E., {Schneider} P., {Semboloni}
  E., {van Waerbeke} L., {Velander} M., 2010, \aap, 516, A63

\bibitem[{{Viola} {et~al.}(2014){Viola}, {Kitching}, \& {Joachimi}}]{viola14}
{Viola} M., {Kitching} T.~D., {Joachimi} B., 2014, \mnras, 439, 1909

\bibitem[{{Whittaker} {et~al.}(2014){Whittaker}, {Brown}, \&
  {Battye}}]{whittaker14}
{Whittaker} L., {Brown} M.~L., {Battye} R.~A., 2014, \mnras, 445, 1836

\bibitem[{{Zuntz} {et~al.}(2013){Zuntz}, {Kacprzak}, {Voigt}, {Hirsch}, {Rowe},
  \& {Bridle}}]{zuntz13}
{Zuntz} J., {Kacprzak} T., {Voigt} L., {Hirsch} M., {Rowe} B., {Bridle} S.,
  2013, \mnras, 434, 1604

\end{thebibliography}

\renewcommand{\theequation}{A-\arabic{equation}}
\renewcommand{\thefigure}{A-\arabic{figure}}
% redefine the command that creates the equation no.
\setcounter{equation}{0}  % reset counter 
\setcounter{figure}{0}
\section*{Appendix A: Residual bias in the iterative method}
\label{ap:res_bias}
We derive the residual bias in the shear estimates as a function of the number of iterations used.

If we assume the zeroth-order shear estimate can be written as equation (\ref{eq:zeroth_bias}), the first-order iteration gives the shear estimate as
\begin{align}\label{eq:first_est_noise}
\hat{g}_i^{(1)}=&\hat{g}_i^{(0)}-\left(\hat{g}_{i,\mathrm{sim}}^{(1)}-\hat{g}_i^{(0)}\right),\nonumber\\
=&\hat{g}_i^{(0)}-m'\hat{g}_i^{(0)}-c_i'+\delta g_i^{(1)}.\nonumber\\
\end{align}
We can substitute equation (\ref{eq:zeroth_bias}) into equation (\ref{eq:first_est_noise}), giving
\begin{equation}\label{eq:first_est_simp}
\hat{g}_i^{(1)}=g_i-m_i'\left(m_i'g+c_i'\right)+\left(1-m'\right)\delta g_i^{(0)}+\delta g_i^{(1)}.
\end{equation}
The second iteration yields
\begin{align}\label{eq:second_est}
\hat{g}_i^{(2)}=&\hat{g}_i^{(0)}-m_i'\hat{g}_i^{(1)}-c_i'+\delta g_i^{(2)},\nonumber\\
=&g_i+m_i'^2\left(m_i'g_i+c_i'\right)+\left[1-m_i'\left(1-m_i'\right)\right]\delta g_i^{(0)}\nonumber\\
&-m_i'\delta g_i^{(1)}+\delta g_i^{(2)}.
\end{align}
From here we see that the error terms propagate linearly through the iterations. Hence, as the mean of these terms is zero, there will be no residual bias contribution from these terms. If we ignore the noise terms, we see that the shear estimate from the $n^{\mathrm{th}}$ iteration can be written as
\begin{equation}\label{eq:nth_est_simp}
\hat{g}_i^{(n)}=g_i+\left(-m_i'\right)^n\left(m_i'g_i+c_i'\right),
\end{equation}
and therefore the bias on the $n^{\mathrm{th}}$ iteration is
\begin{equation}\label{eq:nth_bias_simp}
\left<\hat{g}_i^{(n)}-g_i\right>=\left(-m_i'\right)^n\left(m_i'g_i+c_i'\right).
\end{equation}

\renewcommand{\theequation}{B-\arabic{equation}}
\renewcommand{\thefigure}{B-\arabic{figure}}
% redefine the command that creates the equation no.
\setcounter{equation}{0}  % reset counter 
\setcounter{figure}{0}
\section*{Appendix B: Error on the first iteration}
We discuss the first-order approximation of the error on the angle-only shear estimates with a first-order iteration. For each galaxy in the field, we assume that there is an identical galaxy with the intrinsic ellipticity rotated by $90^{\circ}$.

Let us begin by assuming a first-order approximation of the $F_1\left(\left|\bm{g}\right|\right)$ function (discussed in W14), such that 
\begin{equation}\label{eq:approx_F}
F_1\left(\left|\bm{g}\right|\right)\approx u\left|\bm{g}\right|,
\end{equation}
where $u$ is the first-order coefficient. We write the measured position angle of a galaxy as equation (\ref{eq:alpha_err}) and assume that errors on the position angles are distributed symmetrically about zero. The zeroth-order estimate, $\hat{g}_1^{(0)}$ (with a similar analysis also holding for $\hat{g}_2^{(0)}$), can then be written as an average over the galaxy pairs
\begin{align}\label{eq:est_pairs}
\hat{g}_1^{(0)}\approx & \frac{1}{u}\sum_{i=1}^{\frac{N}{2}}\biggl[\bar{w}_i\left(\alpha_i+2\delta\alpha_i^{(1)}\right)\cos\left(2\alpha_i+2\delta\alpha_i^{(1)}\right)\nonumber\\
&+\bar{w}_i\left(2\alpha_i+\pi+2\delta\alpha_i^{(2)}\right)\cos\left(2\alpha_i+\pi+2\delta\alpha_i^{(2)}\right)\biggr],
\end{align}
where there are $N/2$ galaxy pairs, and where $\delta\alpha_i^{(1)}$ is the error on the position angle of the first galaxy in the pair and $\delta\alpha_i^{(2)}$ is the error on the position angle of the corresponding $90^{\circ}$ rotated galaxy. The weighting function $\bar{w}\left(\alpha_i\right)$ is normalized and given as
\begin{equation}
\bar{w}\left(\alpha_i\right)=\frac{w\left(\alpha_i\right)}{\sum_{i=1}^Nw\left(\alpha_i\right)},
\end{equation}
where the summation is over all galaxies in the field.

From this form of the estimator, we can write an approximate form for the error on the zeroth-order estimate in the limit $\bm{g}\rightarrow0$ as
\begin{align}\label{eq:err_pairs}
\sigma_{\hat{g}_1^{(0)}}^2\approx & \frac{N}{2u^2}\biggl<\biggl[\bar{w}_i^{(1)}\cos\left(2\alpha+2\delta\alpha^{(1)}\right)\nonumber\\
&+\bar{w}_i^{(2)}\cos\left(2\alpha+\pi+2\delta\alpha^{(2)}\right)\biggr]^2\biggr>,\nonumber\\
\approx&\frac{1}{2u^2}\left(\sum_{i=1}^N\left(\bar{w}\left(\alpha_i\right)\right)^2-\frac{1}{N}\beta_{\mathrm{wc}}^2\right),
\end{align}
where $\beta_{\mathrm{wc}}$ is the weighted mean cosine of the error distribution,
\begin{equation}\label{eq:beta_wc}
\beta_{\mathrm{wc}}=\sum_{i=1}^N\bar{w}\left(\hat{\alpha}_i\right)\cos\left(2\delta\alpha_i\right).
\end{equation}
We have assumed that the weighting scheme has successfully removed all contributions from an additive bias. From this approach we find that $\sigma_{\hat{g}_1^{(0)}}\approx\sigma_{\hat{g}_2^{(0)}}\equiv\sigma_{\hat{\bm{g}}^{(0)}}$. A similar approach can be applied to the errors on the shear estimates recovered from the first iteration simulations but with the number of galaxies being dependent on the number of galaxies in the simulations; for our analyses we simulated $N_{\mathrm{sim}}=10^{5}$ galaxies as discussed in Section {\ref{sec:g3_sims}. The error on the first iteration of the shear estimator can then be found by looking at equation (\ref{eq:first_est_simp}) where the error on the zeroth-order shear estimate, $\hat{\bm{g}}^{(0)}$, is $\delta g_1^{(0)}$ (as defined in equation (\ref{eq:zeroth_bias})) and the error on the estimate from the first-order simulations, $\hat{\bm{g}}_{\mathrm{sim}}^{(1)}$, is $\delta g_1^{(1)}$ (as defined in equation (\ref{eq:first_it_clarify})). Assuming that the errors on the shear estimates are Gaussian (which is expected to be true to first-order in the shear due to the central limit theorem), the error on the first-order shear estimate is
\begin{equation}\label{eq:err_est_it}
\sigma_{\bm{g}}^2\approx\left(1-m'\right)^2\sigma_{\hat{\bm{g}}^{(0)}}^2+\sigma_{\hat{\bm{g}}^{(1)}}^2,
\end{equation}
where $\sigma_{\hat{\bm{g}}^{(0)}}^2\equiv\left<\delta g_1^{(0)^2}\right>$ and $\sigma_{\hat{\bm{g}}^{(1)}}^2\equiv\left<\delta g_1^{(1)^2}\right>$. In the absence of an additive bias, the multiplicative bias is effectively due to the $\beta_c$ term in equation (\ref{eq:cov_c_s}). However, when using weighted trigonometric functions $\beta_c=\beta_{\mathrm{wc}}$, such that $m'=\beta_{\mathrm{wc}}-1$. 

\label{lastpage}

\end{document}